\documentstyle[preprint,prl,aps,epsf]{revtex}
\begin{document}
\draft

\title{
Transport of a Luttinger liquid in the presence of a time dependent
impurity
}

\author{
{\bf D. Schmeltzer}\\
Department of Physics\\
City College of the City University of New York\\
Convent Ave at 138 St, New York, NY 10031, USA \footnote{Permanent address}\\
and\\
Department of Physics\\
Technion - Israel Institute of Technology\\
Haifa 32000, Israel\\}

\maketitle
\begin{abstract}
We show that the macroscopic current and charge can be formulated as a
Quantum Mechanical zero mode problem. We find that the current is given
by the velocity operator of a particle restricted to move around a circle.
As an explicit example we investigate a Luttinger liquid of length $L$
which is perturbed by a time dependent impurity. Using the statistical
mechanics of zero modes we computed the non-equilibrium current. In
particular we show that in the low temperature limit, $L_T/L>1$, the
zero mode method introduced here becomes essential for computing the
current.\\
\end{abstract}
 
\newpage
\narrowtext

\section{Introduction} \label{sec-1}

In a two dimensional electron gas, the presence of an external potential
gives rise to a one-dimensional conducting channel named quantum wire
\cite{01}. Quantum wires are characterized by a quantized conductance
of $2e^2/h$ per channel. The presence of the electron-electron
interaction is characterized by the interaction parameter ``$K$", which
affects the tunneling conductance \cite{02,03}. In quantum Hall
experiments, the inter-edge tunneling is also characterized by the
interaction parameter ``$K$" \cite{04}.
The analysis of the transport phenomena requires the understanding of the
macroscopic quantities such as charge and current at zero momentum. We
will show within the Hamiltonian formalism that the quantized zero mode
operators can be used to compute the macroscopic current in the presence 
of interaction and time dependent scatterers.
As an explicit example we will compute the non-equilibrium current for a
Luttinger liquid, perturbed by a time dependent backscattering impurity.
Using the interaction picture we will find an exact expression for the zero
mode coordinates. The time derivative of the zero mode will give the
non-equilibrium current. We will compute the current in two limits,
$L/L_T>1$ - quantum wire case and $L/L_T<1$ - mesoscopic case ($L_T$ is
the thermal coherent length given by $L_T=\hbar v/(K_BT)$).
Comparing with the results given in ref.\cite{02} for $L\rightarrow\infty$
we find similar results for (``$f$" is the impurity frequency and ``$f_{DS}$"
is the equivalent frequency for the drain-source voltage) 
$f=0$, $f_{DS}\not=0$, and $L/L_T>1$.
For the time dependent impurity, $f\not=0$ and $f_{DS}\not=0$ the current
is affected by the frequency $f$ only when $h|f+f_{DS}|>K_BT$.
We find an interesting result in the limit $L_T/L>1$. In this limit the
single particle fermionic spectrum (absent in the standard bosonization)
reproduced by the zero mode becomes essential for the calculation of the
current. We find for $f\not=0$ and $f_{DS}=0$ that the current ``$I$" is
proportional to $ef$ and is periodic in $eV_{GS}$, the static chemical
potential, reflecting the
quantization of charge.\\

The plan of this paper is as the follows: Chapter \ref{sec-2} is devoted
to the method of the zero mode bosonization. In Chapter \ref{sec-3} we
introduce the macroscopic current which is identified with the time
derivative of the macroscopic coordinate. Chapter \ref{sec-4} contains
the analysis of the thermodynamics of the Luttinger liquid. In Chapter
\ref{sec-5} we compute the macroscopic current for the time dependent
impurity potential. Chapter \ref{sec-6} is restricted to the computation
of the conductance for the Luttinger liquid. Finally in Chapter \ref{sec-7}
we conclude the paper.\\

\section{The zero mode bosonization} \label{sec-2}

Bosonization has become the standard method in one dimension since it
allows to incorporate in a trivial way the interactions. For transport
problems one must be able to add and subtract particles, such a thing
is not trivial within the Bosonization method \cite{05}. This problem
has been solved by using zero modes in addition to the
particle-hole bosonic variables \cite{05,06,07,08}. Here we will show how
the zero mode method can be used for computing macroscopic quantities
such as current and charge. Following Ref.\cite{05,06} we replace the
spinless Fermion $\psi(x)$ by
$\psi(x)=e^{iK_Fx}\psi_R(x)+e^{-iK_Fx}\psi_L(x)$ where $R$ and $L$
represent the right and left movers, with $K_F=\pi/2a$,
$\psi_R(x)=\chi_R\tilde{\psi}_R(x)$, and
$\psi_L(x)=\chi_L\tilde{\psi}_L(x)$ where $\chi_i=A_i+A_i^{\dagger}$,
$i=R,L$ such that $\{A_i,A_j^{\dagger}\}=\delta_{i,j}$,
$\{A_i,A_j\}=0$ are the fermionic degrees of freedom (``$x$"
independent) which keep the anticommutation relation between the left
and right movers \cite{07} (An alternative possibility used in Ref.\cite{05}
is to represent the $\chi_i$ in terms of the Klein's factors). The operators
$\tilde{\psi}_R(x)$ and $\tilde{\psi}_L(x)$ are expressed in terms of
bosonic variables and zero modes for a chain of length $L$,

\[
  \tilde{\psi}_R(x)=\sqrt{\frac{K_F}{2\pi}}e^{i\alpha_R}
  e^{i\frac{2\pi}{L}(\hat{N}_R-\frac{1}{2})x}e^{i\theta_R(x)}
\]
\begin{equation}
\label{eq-01}
  \tilde{\psi}_L(x)=\sqrt{\frac{K_F}{2\pi}}e^{-i\alpha_L}
  e^{-i\frac{2\pi}{L}(\hat{N}_L-\frac{1}{2})x}e^{-i\theta_L(x)}
\end{equation}

\noindent The fermionic fields given in Eq.\ref{eq-01} obey antiperiodic
boundary conditions. $\theta_R(x)$ and $\theta_L(x)$ represent the
particle-hole excitation with \underline{no} zero mode which obey
$\theta_{R/L}(x+L)=\theta_{R/L}(x)$.
$\alpha_R$, $\hat{N}_R$ and $\alpha_L$, $\hat{N}_L$ are the zero modes
which obey the commutation rules:
$[\alpha_R,-\hat{N}_R]=[\alpha_L,\hat{N}_L]=i$ \cite{07}.
The operators $\hat{N}_R$ and $\hat{N}_L$ measure the number of fermions,
$\hat{N}_R|N_R;\theta_R\rangle=N_R|N_R;\theta_R\rangle$,
$\hat{N}_L|N_L;\theta_L\rangle=N_L|N_L;\theta_L\rangle$
with the eigen values $N_R$ and $N_L$. $\alpha_R$ and $\alpha_L$ play
the role of the coordinate of a particle on a circle,
$0\le\alpha_{R/L}\le 2\pi$, and $N_R$, $N_L$ are the canonical conjugate
momentum for a rotator. As a result
$e^{\mp i\alpha_R}$ and $e^{\pm i\alpha_L}$
act as a creation (annihilation) operator of particles,
$e^{\mp i\alpha_R}|N_R;\theta_R\rangle=|N_R\pm 1;\theta_R\rangle$,
$e^{\pm i\alpha_L}|N_L;\theta_L\rangle=|N_L\mp 1;\theta_L\rangle$.\\

\section{The macroscopic current} \label{sec-3}

Next we will introduce the macroscopic currents. We will use the method
used in quantum mechanics where one replaces the ``momentum" operator
with the ``velocity" one. In each channel we add $N_R$ and $N_L$ charges.
using the balistic time $f_B^{-1}=L/v_F$ we introduce two currents:

\begin{equation}
\label{eq-02}
   \tilde{I}_R=-\frac{ev_F}{L}N_R;
   \;\;\;
   \tilde{I}_L=\frac{ev_F}{L}N_L;
   \;\;\;
   \tilde{I}=\tilde{I}_R+\tilde{I}_L=\frac{ev_F}{L}(N_L-N_R).
\end{equation}

\noindent Using the fact that the coordinator operator $\alpha_R$ is
conjugate to the momentum operator $\hat{N}_R$ and $\alpha_L$ is
conjugate to $\hat{N}_L$ allows to define the velocity operators,
$\dot{\alpha}_R$ and $\dot{\alpha}_L$, according to the rules of quantum
mechanics. Using the Heisenberg equation of motion, we will identify
the ``velocity'' operator with the current one. We start by defining
the chiral currents $I_L$, $I_R$, and $I=I_R+I_L$,

\begin{equation}
\label{eq-03}
   I_R=\frac{e}{2\pi}\frac{\alpha_{R,H}(t)}{dt};
   \;\;\;
   I_L=\frac{e}{2\pi}\frac{\alpha_{L,H}(t)}{dt};
   \;\;\;
   I=\frac{e}{2\pi}\frac{\alpha_{H}(t)}{dt}.
\end{equation}

\noindent In Eq.(\ref{eq-03}), $\alpha_{R,H}(t)$, $\alpha_{L,H}(t)$,
and $\alpha_{H}(t)$ are the Heisenberg operators:

\begin{equation}
\label{eq-04}
   \alpha_{L/R,H}(t)=e^{iHt/\hbar}\alpha_{L/R}e^{-iHt/\hbar};
   \;\;\;
   \hat{N}_{L/R,H}(t)=\hat{N}_{L/R}
\end{equation}

\noindent with $\alpha=\alpha_L+\alpha_R$ being the coordinate,
$\hat{J}=\hat{N}_L-\hat{N}_R$ the current,
$\hat{Q}=\hat{N}_L+\hat{N}_R$ the charge, and $[\alpha,\hat{J}]=i2$.\\

Using the case of non-interacting zero mode hamiltonian allows to
show the identity between the definitions in Eqs. (\ref{eq-02}) and
(\ref{eq-03}). We consider the hamiltonian:

\[
   H_{L/R}^{(n=0)}=\frac{\pi v_F \hbar}{L}\hat{N}_{L/R}^2
\]
\begin{equation}
\label{eq-05}
   H_0^{(n=0)}=H_{L}^{(n=0)}+H_{R}^{(n=0)}
   =\frac{\pi v_F \hbar}{2L}
   \left(\hat{J}^2+\hat{Q}^2\right).
\end{equation}

\noindent The Heisenberg equation of motion gives:

\begin{equation}
\label{eq-06}
   I_{L/R}=\frac{e}{2\pi}\frac{d\alpha_{L/R,H}}{dt}
   =\frac{e}{i\hbar} [\alpha_{L/R,H},H_{L/R}^{(n=0)}]
   =\pm\frac{ev_F}{L}\hat{N}_{L/R}
\end{equation}

\noindent with

\[
  \hat{N}_{L/R,H}(t)=\hat{N}_{L/R};
  \;\;\;\;\;
  \hat{J}_H(t)=\hat{J}
\]
\begin{equation}
\label{eq-07}
   I=\frac{e}{2\pi}\frac{d\alpha_H}{dt}=\frac{e}{i\hbar}
   [\alpha_H,H_0^{(n=0)}]=\frac{ev_F}{L}\hat{J}_H.
\end{equation}

\noindent Using an eigenstate with a fixed number of fermions,
$|N_R,\theta_R\rangle \bigotimes |N_L,\theta_L\rangle$,
allows to establish that the expectation values 
$\langle I_L\rangle$, $\langle I_R\rangle$, and $\langle I\rangle$
given in Eqs.(\ref{eq-06}) and Eqs.(\ref{eq-07}) are identical
with the definition given in Eq.(\ref{eq-02}).\\

Next we consider the Luttinger liquid zero mode hamiltonian and
show that the current operator is changed such that $v$ is replaced
by $v_J=vK$.

\begin{equation}
\label{eq-08}
   H_0^{(n=0)}=\frac{\pi v \hbar}{2L}
   \left[K\hat{J}^2+\frac{\hat{Q}^2}{K}\right],
   \;\;\;\;\;
   K\le 1.
\end{equation}

\noindent Here the Heisenberg equation of motion gives the result
obtained in Ref.\cite{05},

\begin{equation}
\label{eq-09}
   I=\frac{e}{2\pi}\frac{\alpha_H}{dt}=\frac{e}{i\hbar}
   [\alpha_H,H_0^{(n=0)}]=\frac{evK}{L}\hat{J}_H.
\end{equation}

For the remaining part we will compute the macroscopic zero mode current
$I=[e/(2\pi)]d\alpha_H/dt$ for a Luttinger liquid in the
presence of a time dependent backscattering impurity potential $H_1(t)$
localized at $x=0$.

\begin{equation}
\label{eq-10}
  H=H_0+H_1(t)
\end{equation}
\begin{equation}
\label{eq-11}
  H_0=H_0^{(n=0)}+H_0^{(n\not=0)}
\end{equation}
\begin{equation}
\label{eq-12}
  H_0^{(n=0)}=\frac{\pi v\hbar}{L}
  [\frac{K}{2}\hat{J}^2+\frac{1}{2K}\hat{Q}^2]
\end{equation}
\begin{equation}
\label{eq-13}
  H_0^{(n\not=0)}=\int_0^L\frac{dx}{2\pi} v\hbar
  [\frac{K}{2}(\partial_x\phi)^2+\frac{1}{2K}(\partial_x\theta)^2]
\end{equation}
\begin{equation}
\label{eq-14}
  H_1(t)=\frac{\lambda}{2}[\chi_R \chi_L e^{-i\alpha} e^{-i\theta(x=0)}
  e^{-i\omega t} + \; h.c. \; ], \;\;\; t>0.
\end{equation}

\noindent In obtaining Eqs.\ref{eq-10}-\ref{eq-14} we have used the bosonic
fields $\theta$ and $\phi$:
$\theta(x)=\theta_R+\theta_L$, $\phi(x)=\theta_L-\theta_R$.
Similarly for zero modes we define the charge and current operators:
$\hat{Q}=\hat{N}_R+\hat{N}_L$, $\hat{J}=\hat{N}_L-\hat{N}_R$,
$\alpha=\alpha_L+\alpha_R$, $[\alpha,\hat{J}]=2i$.
The operators $\hat{Q}$ and $\hat{J}$ have the eigenvalues $Q=N_R+N_L$
and $J=N_L-N_R$. $H_0$ represents the Luttinger liquid model for a wire
of length $L$ characterized by the repulsive interaction $K\le 1$ and
Fermi velocity $v$..
$H_1(t)$ represents a time dependent backscattering impurity driven at a
frequency $\omega=2\pi f$ and momentum $q\sim 2K_F$. (Microscopically one
obtains such a term when we couple a driven optical phonon field with
a frequency $\omega=\omega(q)$ and momentum $q\sim 2K_F$ to the
electrons. It is  assumed that the coupling occurs only in a
restricted region in space $d\ll L$ which justifies to approximate the
problem by a localized impurity.) The effect of the time dependent
impurity is similar to an oscillating atom which induces a charge
density wave, $\psi_R^{\dagger}\psi_L\Delta_{CDW} +\;h.c.$,
$\Delta_{CDW}\sim\frac{\lambda}{2}e^{-i\omega t}$. In the absence of any
other source a charge density wave will depend on the relative phase
between the field and the electrons.\\

The macroscopic current for the hamiltonian $H$ given in Eqs.
(\ref{eq-10}) to (\ref{eq-14}) is obtained from the Heisenberg equation
of motion:

\begin{equation}
\label{eq-15}
   I_H(t)=\frac{e}{2\pi}\frac{\alpha_H}{dt}=\frac{e}{i\hbar}
   [\alpha_H,H]=\frac{evK}{L}\hat{J}_H(t).
\end{equation}

\noindent From Eq.(\ref{eq-15}) we observe that the current is
determined by the Heisenberg representation,

\begin{equation}
\label{eq-16}
   \hat{J}_H(t)=\exp\left(\frac{i}{\hbar}Ht\right) \hat{J}
   \exp\left(-\frac{i}{\hbar}Ht\right)
\end{equation}

\noindent For convenience we will use the interaction picture, where

\[
   \alpha_I(t)=\exp\left(\frac{i}{\hbar}H_0^{(n=0)}t\right) \alpha(0)
   \exp\left(-\frac{i}{\hbar}H_0^{(n=0)}t\right);
\]
\begin{equation}
\label{eq-17}
   \hat{J}_I(t)=\exp\left(\frac{i}{\hbar}H_0^{(n=0)}t\right) \hat{J}(0)
   \exp\left(-\frac{i}{\hbar}H_0^{(n=0)}t\right) \equiv \hat{J}
\end{equation}

\noindent and

\begin{equation}
\label{eq-18}
   \alpha_H(t)=U^{\dagger}(t,0)\alpha_I(t)U(t,0);
   \;\;\;
   \hat{J}_H(t)=U^{\dagger}(t,0)\hat{J}_I(t)U(t,0)
\end{equation}

\noindent where $U(t,0)$ is the unitary evolution operator,

\begin{equation}
\label{eq-19}
   U(t,0)=T\exp\left[-\frac{i}{\hbar}\int_0^t dt_1 H_I(t_1)\right];
   \;\;\;
   H_I(t)=\exp\left(\frac{i}{\hbar}H_0t\right) H_1(t)
   \exp\left(-\frac{i}{\hbar}H_0t\right).
\end{equation}

\noindent Combining Eqs. (\ref{eq-16}), (\ref{eq-17}), and (\ref{eq-18}),
we obtain the Heisenberg representation of the current $I_H(t)$,

\begin{equation}
\label{eq-20}
   I_H(t)=\frac{evK}{L} U^{\dagger}(t,0) \hat{J} U(t,0)
\end{equation}

\noindent In order to compute the current we have to perform the
statistical average of Eq.(\ref{eq-20}). In particular we will be
interested in computing the current $I_H(t)$ in the presence of an
external source of voltage $\tilde{V}_{DS}$. This requires to add to the zero
mode hamiltonian a source term of the form $e\tilde{V}_{DS}\hat{J}$. As a
result, the zero mode hamiltonian $H_0^{(n=0)}$ is replaced by
$H_0^{(n=0)}+e\tilde{V}_{DS}\hat{J}$, causing a change in the current
operator $I_H(t)\rightarrow I_H(t)+(e^2/h)\tilde{V}_{DS}$. This simple
procedure
has been questioned in Ref.\cite{09} where the external voltage source
connected to the wire is screened. As a result it has been argued that
the screened voltage source must be used and not the applied voltage
$\tilde{V}_{DS}$. In this paper we will adopt a thermodynamic point of view.
We will replace the external voltage $\tilde{V}_{DS}$ by two chemical
potentials $\mu_L$ and $\mu_R$. As a result no external source will be
added to the zero mode Hamiltonian $H_0^{(n=0)}$. Therefore the Heisenberg
equation of motion for the current will be given by Eq.(\ref{eq-20}).
The only change will occur at the level of the thermodynamic expectation
values which will depend on the chemical potentials in the reservoirs.
(The chemical potentials in the reservoirs are different from the ones
in the wires. This difference reflects the screening phenomena.) In this
paper we will define the conductance with respect to the chemical potential
in the reservoir. Following Ref.\cite{10} we will consider
that the thermal reservoir has two chemical potentials, $\mu_R$ and
$\mu_L$. We will choose $(\mu_L+\mu_R)/2\equiv eV_{GS}$ and
$\mu_L-\mu_R\equiv eV_{DS}$. The presence of the chemical potential
has no effects on the equation of motion. Only the statistical average
will be affected. The reservoir will be characterized by the partition
function, $Z=Z^{(n\not=0)}Z^{(n=0)}$,

\begin{equation}
\label{eq-21}
   Z^{(n\not=0)}=\;Tr\; e^{-\beta H_0^{(n\not=0)}}
\end{equation}
\begin{equation} 
\label{eq-22}
   Z^{(n=0)}=\;Tr\; \left[ e^{-\beta H_0^{(n=0)}}
    e^{-\beta(\mu_R\hat{N}_R+\mu_L\hat{N}_L)} \right]
\end{equation}
 
\noindent In order to compute the non-equilibrium current caused by
$H_1(t)$, we will compute the statistical average using the partition
function $Z$. We introduce the notation:

\[
  \rho(H_0)=\rho\left(H_0^{(n\not=0)}\right)\rho\left(H_0^{(n=0)}\right)\equiv
  \langle\langle\cdots\rangle\rangle
\]

\noindent where

\begin{equation}  
\label{eq-23} 
   \rho\left(H_0^{(n\not=0)}\right)=\left[ Z^{(n\not=0)} \right]^{-1}
   e^{-\beta H_0^{(n\not=0)}} \equiv \langle \cdots \rangle_{n\not=0};
\end{equation} 
\begin{equation}
\label{eq-24}
   \rho\left(H_0^{(n=0)}\right)=\left[ Z^{(n=0)} \right]^{-1}
   e^{-\beta H_0^{(n=0)}} e^{-\beta(\mu_R\hat{N}_R+\mu_L\hat{N}_L)}
   \equiv \langle \cdots \rangle_{n=0}.
\end{equation}

\section{Thermodynamics of the zero mode Luttinger liquid} \label{sec-4}

We will compute the zero mode partition function
$Z^{(n=0)}\equiv Z^{(n=0)}(\mu_R,\mu_L)$. Using the partition function
we will compute the zero mode expectation values,
$\langle\hat{N}_R\rangle_{n=0}$, $\langle\hat{N}_L\rangle_{n=0}$
$\langle\hat{N}_R^2\rangle_{n=0}$, and $\langle\hat{N}_L^2\rangle_{n=0}$.
These result will be used to compute the current in the presence of the
time dependent perturbation $H(t_1)$ given by Eq.(\ref{eq-14}). We will
consider two cases, $L/L_T \gg 1$ and $L/L_T \le 1$.\\

\noindent {\bf a)} The $L/L_T \gg 1$ case. In this limit the sum over
$N_R$ and $N_L$ in Eq.(\ref{eq-22}) is replaced by an Gaussian integral.
As a result we obtain:

\[
   \langle\hat{N}_R\rangle_{n=0} \longrightarrow
   \frac{\mu_R}{h}\frac{L}{Kv},
   \;\;\;\;\;
   \langle\hat{N}_L\rangle_{n=0} \longrightarrow
   \frac{\mu_L}{h}\frac{L}{Kv},
\]
\begin{equation} 
\label{eq-25}
   \langle\hat{J}\rangle_{n=0} \longrightarrow
   \frac{eV_{DS}}{h}\frac{L}{Kv}, 
   \;\;\;\;\; 
   \langle\hat{Q}\rangle_{n=0} \longrightarrow 
   2 \frac{eV_{GS}}{h}\frac{L}{Kv}
\end{equation} 
  
\noindent and

\[
   \langle\hat{N}_R^2\rangle_{n=0}-\langle\hat{N}_R\rangle_{n=0}^2
   =\langle\hat{N}_L^2\rangle_{n=0}-\langle\hat{N}_L\rangle_{n=0}^2
   \longrightarrow \frac{1}{2\pi K} \left(\frac{L}{L_T}\right)
\]
\begin{equation} 
\label{eq-26}
   \langle\hat{J}^2\rangle_{n=0}-\langle\hat{J}\rangle_{n=0}^2 
   =\langle\hat{Q}^2\rangle_{n=0}-\langle\hat{Q}\rangle_{n=0}^2 
   \longrightarrow \frac{1}{\pi K} \left(\frac{L}{L_T}\right).
\end{equation}

\noindent {\bf b)} The $L/L_T \le 1$ case. This represents the low
temperature limit where single particle excitations are reproduced by the
zero mode theory. For this case standard bosonization is not applicable
since the fermionic distribution function is absent. The partition
function $Z^{(n=0)}(\mu_R,\mu_L)$ is given by:

\begin{equation}  
\label{eq-27}
   Z^{(n=0)}(\mu_R,\mu_L)=\sum_{N_L=-\infty}^{\infty}
   W^{N_L^2}V_L^{N_L} \sum_{N_R=-\infty}^{\infty}
    W^{N_R^2}V_R^{N_R(1+rN_L)}
\end{equation}   
   
\noindent where

\begin{equation}   
\label{eq-28}
   W=\exp\left[-\frac{\pi L_T}{2L}\left(K+K^{-1}\right)\right];
\end{equation}
\begin{equation}    
\label{eq-29} 
   V_L=\exp\left(\frac{\mu_L}{K_BT}\right),
   \;\;\;\;\;
   V_R=\exp\left(\frac{\mu_R}{K_BT}\right);
\end{equation}
\begin{equation}    
\label{eq-30} 
   r=\frac{\pi v \hbar}{L\mu_R}\left(K^{-1}-K\right).
\end{equation}

\noindent For $K=1$, $r=0$ and the Luttinger term $N_LN_R$ is absent
in Eq.(\ref{eq-27}). As a result,
$Z^{(n=0)}(\mu_R,\mu_L)=Z^{(n=0)}(\mu_R)Z^{(n=0)}(\mu_L)$. Using the
Jacobi identity we find:

\begin{equation}
\label{eq-31}
   Z^{(n=0)}(\mu_{L/R})=\prod_{n=1}^{\infty} \left[
   \frac{\left(1+W^{2n-1}V_{L/R}\right)\left(1+W^{2n-1}
   V_{L/R^{-1}}\right)}{\left(1+W^{2n-1}\right)^2}\right].
\end{equation}
 
\noindent Using Eq.(\ref{eq-31}) we find

\begin{equation}
\label{eq-32}
   \langle\hat{N}_{L/R}\rangle_{n=0}=K_BT
   \frac{\partial\ln Z^{(n=0)}(\mu_{L/R})}{\partial \mu_{L/R}}
   =\sum_{n=1}^{\infty}
   \left[1+\exp\beta(\epsilon_n-\mu_{L/R})\right]^{-1}
\end{equation} 
  
\noindent where $\epsilon_n=(2\pi/L)v\hbar(n-1/2)$ is the single particle
spectrum. Similarly we obtain:

\begin{equation}
\label{eq-33}
   \langle\hat{N}_{L/R}^2\rangle_{n=0}-\langle\hat{N}_{L/R}\rangle_{n=0}^2
   =K_BT\frac{\partial\langle\hat{N}_{L/R}\rangle_{n=0}}{\partial\mu_{L/R}}
   =\frac{1}{4}\sum_{n=1}^{\infty}\cosh^{-2}\left(
   \frac{\epsilon_n-\mu_{L/R}}{2K_BT}\right).
\end{equation}  
   
\noindent Next we consider the Luttinger case where $K<1$ and $r\not=0$
in Eq.(\ref{eq-27}). For Eq.(\ref{eq-27}) we perform first the sum over
$N_R$. As a result we find:

\begin{equation} 
\label{eq-34}
   \sum_{N_R=-\infty}^{\infty} W^{N_R^2}V_R^{N_R}V_R^{rN_RN_L}
   \sim V_R^{P(N_L,r)}.
\end{equation}   
    
\noindent $P(N_L,r)$ is a polynomial in $N_L$ which can be obtained in
the limit $L/L_T\ll 1$. We find

\begin{equation}  
\label{eq-35}
   P(N_L,r)\sim \frac{\mu_RL}{\pi v\hbar} \frac{2}{K+K^{-1}}
   \left[ rN_L+\frac{1}{2}(rN_L)^2\right].
\end{equation}   
    
\noindent Substituting Eq.(\ref{eq-34}) into Eq.(\ref{eq-27}) gives:

\begin{equation}   
\label{eq-36}
   \langle\hat{N}_{L/R}\rangle_{n=0} = \sum_{n=1}^{\infty}
   \left[1+\exp\frac{\hat{\epsilon}_n-\mu_{L/R}}{K_BT^*} \right]^{-1}
\end{equation}    
     
\noindent where $T^*$ and $\hat{\epsilon}_n$ are given by:

\[
   T^*\equiv T[1+g(K)]^{-1} ;
   \;\;\;\;\;
   g(K)\equiv 2\left( \frac{K^{-1}-K}{K^{-1}+K} \right),
\]
\begin{equation}    
\label{37}
   \hat{\epsilon}_n \equiv \epsilon_n \left(\frac{K^{-1}+K}{2}\right)
   \left[\frac{1-2\left(\frac{K^{-1}-K}{K^{-1}+K}\right)}{
   1+2\left(\frac{K^{-1}-K}{K^{-1}+K}\right)} \right],
\end{equation}    
     
\noindent and

\begin{equation}     
\label{eq-38}
   \langle\hat{N}_{L/R}^2\rangle_{n=0}-\langle\hat{N}_{L/R}\rangle_{n=0}^2
   =\frac{1}{8}\sum_{n=1}^{\infty}\cosh^{-2}
   \frac{\hat{\epsilon}_n-\mu_{L/R}}{2K_BT^*}
\end{equation}     

\noindent Comparing the case $K\not=1$ to $K=1$ we find that the form is
preserved given that we replace $\epsilon_n$ by $\hat{\epsilon}_n$ and
the temperature $T$ by $T^*$.\\

\section{Computation of the macroscopic current in the presence of a weak
impurity potential} \label{sec-5}

Our starting point will be Eq.(\ref{eq-20}) with $U(t,0)$ controlled by
the impurity hamiltonian $H_1(t)$. For the static impurity case, standard
RG calculation shows that for $K<1$ the stable fixed point of $\lambda$ is
$\lambda^*=\infty$. This means that the perturbation theory around the
unstable fixed point $\lambda^*=0$ must break for length scale
$\ell>\ell_0$, where $\ell_0$ satisfies
$\hat{\lambda}(\ell_0)\simeq\hat{\lambda}\ell_0^{1-K}\simeq 1$. In the
remaining part we will assume that for a finite system of length $L=\ell_0a$,
$\hat{\lambda}\ll 1$, and $\hat{\lambda}(\ell_0)<1$. Therefore for this range
of parameters we can expand $U(t,0)$ in powers of
$\hat{\lambda}\equiv \lambda/\Lambda$ and find

\[
  \langle\langle I_H(t) \rangle\rangle
  =\frac{e}{L} vK \langle\hat{J}\rangle_{n=0}
  +\frac{e}{L}vK(\frac{i}{\hbar})\int_0^tdt_1
  \langle\langle [H_I(t_1),\hat{J}] \rangle\rangle
\]
\begin{equation}
\label{eq-39}
  +\frac{e}{L}vK(\frac{i}{\hbar})^2 \int_{c_t} dt_1 \int_{c_t} dt_2
  \langle\langle T[H_I(t_1)H_I(t_2)] \hat{J} \rangle\rangle+ \cdots.
\end{equation}
 
\noindent In Eq.(\ref{eq-39}), $c_t$ stays for the Keldysh contour:
$0\rightarrow t$, $t\rightarrow 0$, and $0\rightarrow -i\beta$.
The term $\langle\cdots\rangle_{n=0}$ represents the
expectation values with respects to the zero mode Hamiltonian
$H_0^{(n=0)}$. The second term vanishes since it contains
the operators $e^{\pm i\alpha}$ (The expectation value for the terms of
the forms $\langle N_R | e^{\pm i\alpha_R} | N_R \rangle$ and
$\langle N_L | e^{\pm i\alpha_L} | N_L \rangle$ are zero).
The third term decouples into two expectation values,
$\langle\cdots\rangle_{n\not=0}$ (the bosonic part) and
$\langle\cdots\rangle_{n=0}$ (the zero mode part) which is not zero,
since the zero mode coordinate $\alpha$ cancels. As a result we find:
 
\[
  \langle\langle I_H(t)\rangle\rangle
  =\frac{e}{L}vK\langle\hat{J}\rangle_{n=0}
  -e\left(\frac{\lambda}{2\hbar}\right)^2 \frac{2vK}{L}
  \int_0^t dt_1 e^{i(\frac{4\pi}{L}vK) t_1} \cdot
\]
\[
  \int_0^t ds \langle e^{\pm i[\theta_I(s)-\theta_I(0)]}
  \rangle_{n\not= 0} e^{-i \frac{4\pi}{L}vKs}
  \langle\cos\left[\left(\frac{2\pi}{L}vK\hat{J}+\omega\right)s\right]
  \hat{J} \rangle_{n=0}
\]
\begin{equation}
\label{eq-40}
  \stackrel{t\rightarrow\infty}{\leadsto}
  \frac{e}{L}vK\langle\hat{J}\rangle_{n=0}
  -\frac{e}{2(4\pi)}\left(\frac{\lambda}{\hbar}\right)^2
  \int_0^t ds \; F(s) \sin\left(\frac{4\pi}{L} vKs\right)
  \langle \cos\left[\left(\frac{2\pi}{L}vK\hat{J}+\omega\right)s\right]
  \hat{J} \rangle_{n=0}.
\end{equation}

\noindent In Eq.(\ref{eq-40}) we observe that the current is determined by
the zero mode expectation value $\langle\cdots\rangle_{n=0}$ and
separately by the bosonic part
$F(s)\stackrel{\rm def}{=}\langle e^{\pm i[\theta_I(s)-\theta_I(0)]}
\rangle_{n\not=0}$, where
$e^{\pm i\theta_I(t)}=e^{\frac{i}{\hbar}H_0^{(n\not=0)}t}
e^{\pm i\theta(0)} e^{-\frac{i}{\hbar}H_0^{(n\not=0)}t}$
In order to evaluate the current in Eq.(\ref{eq-40}) we have to compute
the zero mode part, $\langle\hat{J}\rangle_{n=0}$ and
$\langle\cos[2\pi(\frac{vK}{L}\hat{J}+f)s]\hat{J} \rangle_{n=0}$.\\

\noindent {\bf a)} We consider first the quantum wire limit
$L/L_T\ll 1$ where $L_T=\hbar v\beta =\frac{\hbar v}{k_B T}$
is the thermal length. Using the results given in Eqs.(\ref{eq-25}) and
(\ref{eq-26}) we obtain:
 
\[
  \langle\hat{J}\rangle_{n=0}\longrightarrow\frac{eV_{DS}}{h}\frac{L}{Kv},
  \;\;\;\;\;
  \frac{eV_{DS}}{h}\stackrel{\rm def}{=}f_{DS},
\]
\begin{equation}
\label{eq-41}
  \langle\cos\left[2\pi\left(\frac{vK}{L}\hat{J}+f\right)s\right]\hat{J}
  \rangle_{n=0}
  \longrightarrow f_{DS}\left(\frac{L}{Kv}\right) \cos[2\pi(f_{DS}+f)s]
\end{equation}

\noindent In Eq.(\ref{eq-41}) $f_{DS}$ is either positive or negative since
it depends on the drain source voltage $V_{DS}$. We introduce the
dimensionless length
``$\ell"$: $\ell\stackrel{\rm def}{=} \frac{L_T}{a}=\ell_T$ for
$|f_{DS}+f|^{-1}>\frac{L_T}{v}$ or
$\ell\stackrel{\rm def}{=} \frac{v}{|f_{DS}+f|a} = \ell_f$ for
$|f_{DS}+f|^{-1} < \frac{L_T}{a}$.
Using the dimensionless length $\ell$ defined in Eq.(\ref{eq-41}) and
the function $F(s)\sim[(2\pi)/(v\Lambda s)]^{2K}$, $s<\hbar\beta$, we find:

\begin{equation}
\label{eq-42}
  I\sim e\;f_{DS}\left[1-\frac{\hat{\lambda}^2c_0}{4(1-K)}\ell^{2(1-K)}\right],
  \;\;\;
  K<1, \;\;\;
  c_0\simeq 1
\end{equation}
 
\noindent From Eq.(\ref{eq-42}) we observe that the current
grows with the increases of temperature (decreases of $\ell$). The results
obtained in Eq.(\ref{eq-42}) are in agreement with the one
obtained in ref.\cite{02}. We want to emphasize that according to our
discussion at the beginning of this section, Eq.(\ref{eq-42}) is valid
only for ``$\hat{\lambda}$" which obeys $\hat{\lambda}\ell^{1-K}<1$.
(For $\ell\rightarrow\infty$ we have to perform a duality transformation
where $\hat{\lambda}\rightarrow\hat{t}=1/\hat{\lambda}$,
$\theta\rightarrow\phi$, and $K\rightarrow\eta=1/K$.)\\

We want to mention that the formalism presented here is applicable
to the ``persistent current" problem \cite{05,12,13,06}. In
particular we want to emphasize that the result of the
``persistent current" follows from Eq.(\ref{eq-41}) once we substitute
$eV_{DS}/h=(2\pi/L)(vK)(\Phi/\Phi_0)$, where ``$\Phi$" is the
external flux and ``$\Phi_0$" is the flux quantum. This correspondence
follows from the fact that in the presence of an external flux ``$\Phi$",
one obtains an equivalent problem with twisted boundary conditions.
From Eq.\ref{eq-01} it follows that the twist of the boundary
condition is realized when one shift the zero mode operator $\hat{J}$
to $\hat{J}+2\Phi/\Phi_0$. As a result the zero mode Hamiltonian
$H_0^{(n=0)}$ will be shifted by,
$(2\pi/L)(vK)\hbar\hat{J}\Phi/\Phi_0+\;Const$. As an explicite example we
consider the effect of the static impurity on the ``persistent current"
in the limit $L/L_T >1$. Using the results given in Eqs.(\ref{eq-41}) and
(\ref{eq-42}), we find in the limit $\Phi/\Phi_0\rightarrow 0$
 
\begin{equation}
\label{eq-43}
  I=\frac{2\pi(vK)}{L}\left(\frac{\Phi}{\Phi_0}\right)
  \left[1-\frac{\hat{\lambda}^2c_0}{4(1-K)}\ell_T^{2(1-K)}\right].
\end{equation}
 
Next we return to the time dependent impurity driven by the frequency
$2\pi f$ when $f_{DS}=0$. We compute the
expectation values $\langle\hat{J}\rangle_{n=0}$,
$\langle\hat{J}^2\rangle_{n=0}$ and
$\langle\cos[2\pi(\frac{vK}{L}\hat{J}+f)s]\hat{J}\rangle_{n=0}$.
We find from Eqs. (\ref{eq-25}) and (\ref{eq-26}):

\begin{equation}
\label{eq-44}
  \langle\hat{J}\rangle_{n=0}=0,
  \;\;\;\;\;
  \langle\hat{J}^2\rangle_{n=0}=\frac{1}{\pi K}\left(\frac{L}{L_T}\right)
\end{equation}
\begin{equation}
\label{eq-45}
  \langle\cos\left[2\pi\left(\frac{vK}{L}\hat{J}+f\right)s\right]\hat{J}
  \rangle_{n=0}
  =-\sin(2\pi f\;s)\langle\sin\left(\frac{2\pi}{L}vK\hat{J}s\right)\hat{J}
  \rangle_{n=0}
\end{equation}

\noindent As a result we obtain:

\begin{equation}
\label{eq-46}
  I\simeq ef\left[\frac{\hat{\lambda}^2}{4(2-K)}\left(\frac{L_T}{L}\right)
  \ell^{2(1-K)}\right],
  \;\;\;
  K<1.
\end{equation}

\noindent Eq.(\ref{eq-46}) shows that in the quantum wire limit
$\frac{L_T}{L}\rightarrow 0$, $I\rightarrow 0$. The current in
Eq.(\ref{eq-46}) is induced by the driven field
$\Delta_{CDW}\sim\frac{\lambda}{2}e^{-i\omega t}$. The presence of this field
shifts the value of $\alpha$ to $\alpha+\omega t$.
As a result the current will depend on the
frequency ``$f$". The current will be non-zero if $2\pi f\frac{L_T}{v}<1$.
(For high frequencies $2\pi f\frac{L_T}{v}\gg 1$ the current will vanish.)
The direction of the current is determined by the initial phase of the
driven field. Formally the direction of the current is determined by the
total phase
$\alpha_I(t)+\omega t=\alpha(0)+\frac{2\pi vK}{L}\hat{J}t+\omega t$.
When the phase increases, the direction of the current is positive.
The positive current ($\hat{J}>0$) corresponds to $\mu_L-\mu_R>0$.
Therefore a positive current corresponds to the direction of a current
which will flow when $\mu_L>\mu_R$.\\

\noindent {\bf b)} Next we consider the low temperature limit
$L_T/L>1$. For this case we will make use of the results obtained in
Eqs. (\ref{eq-36}) to (\ref{eq-38}). For simplicity, we will consider
two problems: {\bf 1)} $\mu_L=\mu_R=\mu$ and $f\not=0$;
{\bf 2)} $\mu_L-\mu_R=eV_{DS}\not= 0$ and $f=0$.\\

We consider first problem {\bf 1)}. According to Eqs.
(\ref{eq-36})-(\ref{eq-38}) we find:

\[
   \langle\hat{J}\rangle_{n=0}=0,
   \;\;\;
   \langle\hat{Q}\rangle_{n=0}=2\sum_{n=1}^{\infty}
   \left[ 1+\exp\frac{\hat{\epsilon}_n-\mu}{K_BT^*} \right],
\]
\begin{equation}
\label{eq-47}
   \langle\hat{J}^2\rangle_{n=0}=\frac{1}{4}\sum_{n=1}^{\infty}
   \cosh^{-2}\frac{\hat{\epsilon}_n-\mu}{2K_BT^*}.
\end{equation}

\noindent As a result we find from Eq.(\ref{eq-41}) that the current
is given by:

\begin{equation} 
\label{eq-48}
   I \simeq ef\left(\frac{\tilde{f}_B^2}{\tilde{f}_B^2-f^2}\right)
   \hat{\lambda}^2\ell^{2(1-K)}
   \left(\frac{1}{4}\sum_{n=1}^{\infty}
   \cosh^{-2}\frac{\hat{\epsilon}_n-\mu}{2K_BT^*} \right)
\end{equation}
 
\noindent where $\tilde{f}_B=f_BK$ is the ballistic frequency and
$\ell=L/a$. The result in Eq.(\ref{eq-48}) is obtained in the limit
$\tilde{f}_B>f$. At resonance we introduce a broadening $\Delta$ and
find for $|f-\tilde{f}_B|<\Delta$ the result

\begin{equation}  
\label{eq-49}
   I \simeq ef\hat{\lambda}^2\ell^{2(1-K)}
   \left(\frac{1}{4}\sum_{n=1}^{\infty} 
   \cosh^{-2}\frac{\hat{\epsilon}_n-\mu}{2K_BT^*} \right).
\end{equation} 
  
\noindent Eq.(\ref{eq-49}) is valid for $\hat{\lambda}\ell^{(1-K)}\le 1$.
Eqs.(\ref{eq-47})-(\ref{eq-49}) show that the current is periodic each
time the energy level $\hat{\epsilon}_n$ coincides with the charging
potential $V_{GS}=(\mu_L+\mu_R)/(2e)=\mu/e$. In the limit $T\rightarrow 0$
Eq.(\ref{eq-49}) is replaced by a Kronecker-Delta function. Tuning the
length ``$L$" such that $\hat{\lambda}\ell^{1-K}\simeq 1$ we find,

\begin{equation}   
\label{eq-50}
   I \simeq ef \sum_{n=1}^{\infty} \delta(\hat{\epsilon}_n-eV_{GS}).
\end{equation}  
   
\noindent  This result might be relevant to the surface acoustic wave
experiment described in Ref.\cite{14}.\\

In the last part we consider problem {\bf 2)}, the static impurity case.
We have , from Eq.(\ref{eq-36}),

\begin{equation}   
\label{eq-51} 
   \langle\hat{J}\rangle_{n=0}= \sum_{n=1}^{\infty}
   \left( 1+\exp\frac{\hat{\epsilon}_n-\mu_L}{2K_BT^*} \right)^{-1}
   -\sum_{n=1}^{\infty} 
   \left( 1+\exp\frac{\hat{\epsilon}_n-\mu_R}{2K_BT^*} \right)^{-1}.
\end{equation}   
    
\noindent We substitute in $\mu_{L/R}=eV_{GS}\pm eV_{DS}/2$ and find:

\begin{equation}    
\label{eq-52} 
   \langle\hat{J}\rangle_{n=0}= \frac{eV_{DS}}{K_BT^*}
   \left(\frac{1}{4}\sum_{n=1}^{\infty} 
   \cosh^{-2}\frac{\hat{\epsilon}_n-\mu}{2K_BT^*} \right);
   \;\;\;
   \mu\equiv eV_{GS}
\end{equation} 
  
\noindent As a result we find from Eq.(\ref{eq-49}) that for
$\hat{\lambda}\ell^{(1-K)} \equiv \lambda(L/a)^{1-K}<1$,
the current is given by:

\begin{equation}     
\label{eq-53}
   I \simeq K\frac{e^2}{h} V_{DS} \left(\frac{L_T^*}{L} \right)
   \left(\frac{1}{4}\sum_{n=1}^{\infty}  
   \cosh^{-2}\frac{\hat{\epsilon}_n-\mu}{2K_BT^*} \right)
   \left[ 1-\hat{\lambda}^2\ell^{2(1-K)} \;Const.\; \right]
\end{equation}  
   
\noindent where $L_T^*$ is given by $L_T^*=(\hbar v)/(K_BT^*)$.\\

\section{The Luttinger liquid conductance} \label{sec-6}

Using the results given in Sections \ref{sec-4} and \ref{sec-5}
we will compute the conductance for $\hat{\lambda}=0$ and $K\le 1$.
We find that the current is given by:

\begin{equation}      
\label{eq-54}
   I=\frac{e}{L}vK \langle\hat{J}\rangle_{n=0}.
\end{equation}   
    
\noindent  Using the expectation value of $\langle\hat{J}\rangle_{n=0}$
for $\mu_L-\mu_R=eV_{DS}$ we find:\\

\noindent {\bf a)} In the limit $L/L_T\rightarrow\infty$, Eq.(\ref{eq-25})
gives $\langle\hat{J}\rangle_{n=0}=(eV_{DS})/h \cdot L/(Kv)$.
Substituting this value into Eq.(\ref{eq-54}) gives: $I=(e^2/h)V_{DS}$,
therefore $G=e^2/h$.\\

\noindent {\bf b)} In the limit $L/L_T \le 1$, $\langle\hat{J}\rangle_{n=0}$
is given by Eq.(\ref{eq-52}). As a result the conductance is given by:

\begin{equation}       
\label{eq-55}
   G= \frac{e^2}{h} K \left(\frac{L_T^*}{L} \right)
   \left(\frac{1}{4}\sum_{n=1}^{\infty}   
   \cosh^{-2}\frac{\hat{\epsilon}_n-\mu}{2K_BT^*} \right).
\end{equation}

\section{Conclusion} \label{sec-7}

To conclude we can say that the zero mode formalism is the proper one
for computing the macroscopic current. We emphasize that the zero mode
method is essential for obtaining correct results in the limit $L_T/L\ge1$.
In the limit $L_T/L\ge 1$ the current is
controlled by the fermionic spectrum which is completely absent in the
standard bosonization method. As an explicit demonstration of the
method we have computed the Luttinger liquid current in the presence of a
time dependent impurity.\\

\end{document}